\documentclass[aps,prl,twocolumn,amsmath,amssymb,amsfonts,nofootinbib,long,floatfix,showpacs]{revtex4}
\usepackage{epsfig,latexsym,bm,graphicx,epstopdf}

\newcommand\eV{\mbox{eV}}

\newcommand\MeV{\mbox{MeV}}
\newcommand\GeV{\mbox{GeV}}
\newcommand\pc{\mbox{pc}}

\newcommand\Mpc{\mbox{Mpc}}

\newcommand\km{\mbox{km}}
\newcommand\s{\mbox{s}}
\newcommand\G{\mbox{G}}
\newcommand\A{\mathbf{A}}
\newcommand\B{\mathbf{B}}
\newcommand\E{\mathbf{E}}
\newcommand\x{\mathbf{x}}

\newcommand\kk{\mathbf{k}}
\newcommand\ee{{\boldsymbol \varepsilon}}
\newcommand\mPl{m_{\rm Pl}}
\newcommand\C{{\boldsymbol {\mathcal{C}}}}
\newcommand\BB{{\boldsymbol {\mathcal{B}}}}

\begin{document}

\title{SUPERHORIZON MAGNETIC FIELDS}

\author{Leonardo Campanelli$^{1}$}
\email{leonardo.campanelli@ba.infn.it}
\affiliation{$^1$Dipartimento di Fisica, Universit\`{a} di Bari, I-70126 Bari, Italy}

\date{\today}

%***********************************   Abstract   ********************************************%

\begin{abstract}
We analyze the evolution of superhorizon-scale
magnetic fields from the end of inflation till today. Whatever is the mechanism responsible for their generation during inflation, we find that a given magnetic mode with wave number $k$ evolves, after inflation, according to the values of $k\eta_e$, $n_\kk$, and $\Omega_k$, where $\eta_e$ is the conformal time at the end of inflation, $n_\kk$ is the number density spectrum of inflation-produced photons, and $\Omega_k$ is the
phase difference between the two Bogolubov coefficients which characterize the state of that mode at the end of inflation. For any realistic inflationary magnetogenesis scenario, we find that $n_\kk^{-1} \ll |k\eta_e| \ll 1$, and three evolutionary scenarios are possible:
($i$) $|\Omega_k \mp \pi| = \mathcal{O}(1)$, in which case the evolution of the magnetic spectrum $B_k(\eta)$ is adiabatic, $a^2B_k(\eta) = \mbox{const}$, with $a$ being the expansion parameter;
($ii$) $|\Omega_k \mp \pi| \ll |k\eta_e|$, in which case the evolution is superadiabatic,
$a^2B_k(\eta) \propto \eta$;
($iii$) $|k\eta_e| \ll |\Omega_k \mp \pi| \ll 1$ or $|k\eta_e| \sim |\Omega_k \mp \pi| \ll 1$, in which case an early phase of adiabatic evolution is followed, after a time $\eta_\star \sim |\Omega_k \mp \pi|/k$, by a superadiabatic evolution. Once a given mode reenters the horizon, it remains frozen into the plasma and then evolves adiabatically till today. As a corollary of our results, we find that inflation-generated magnetic fields evolve adiabatically on all scales and for all times in conformal-invariant free Maxwell theory, while they evolve superadiabatically after inflation on superhorizon scales in the nonconformal-invariant Ratra model, where the inflaton is kinematically coupled to the electromagnetic field. The latter result supports and, somehow, clarifies our recent claim that the Ratra model can account for the presence of cosmic magnetic fields without suffering from both backreaction and strong-coupling problems.
\end{abstract}

%*********************************************************************************************%


%98.80.-k -> Cosmology
%98.62.En -> Electric and magnetic fields

\maketitle

%*********************************************************************************************%

\section{I. Introduction}

Recent astrophysical observations of gamma-rays spectra of
distant blazars~\cite{Neronov-Vovk,Tavecchio1,Tavecchio2,Caprini0}
strongly indicate the presence of large-scale magnetic fields in cosmic voids.
Together with the ubiquitous presence of large-scale, coherent magnetic fields in clusters of galaxies
and the magnetization of galaxies at both low and high redshifts,
this fact strongly points toward ($i$) the existence of a ``cosmic magnetic field''
that pervades the entire observable universe, ($ii$) that its origin
is primordial (namely, it took place before large-scale structures formation), and ($iii$) it
is a relic of inflation (for reviews on primordial magnetic fields, see~\cite{Widrow,Giovannini,Tsagas,Widrow2,Durrer,Subra}).

However, because of the conformal invariance of classical Maxwell electromagnetism in a
Friedmann-Robertson-Walker background,
photons cannot be created during inflation,
as a consequence of the ``Parker theorem''~\cite{Birrell-Davies,Parker-Toms}.
Quantum effects, nevertheless, can break such a conformal invariance allowing for a generation of
strong magnetic fields~\cite{Dolgov,Campanelli}, or
electromagnetic vacuum fluctuations can either survive and be amplified in marginally
open universes~\cite{Tsagas-Kandus,Barrow-Tsagas,Barrow-et al},
or be ``boosted'' by gravitational waves~\cite{TsagasGW}.

Another possibility to create seed magnetic fields during inflation is to
consider nonstandard, nonconformal-invariant electromagnetic theories.
Starting from the seminal papers by
Turner and Widrow~\cite{Turner-Widrow}, who studied nonconformal couplings between photons and gravity,
and by Ratra~\cite{Ratra}, who considered a conformal-breaking coupling between the inflaton
and the electromagnetic field,
there have been many attempts in this direction
(see, e.g.,~\cite{G1,G2,G3,G4,G5,G6,G7,G8,G9,G10,G11,G12,G13,G14,G15,
G16,G17,G18,G19,G20,G21,G22,G23,G24,G25,G26,G27,G28,G29,G30,G31,G32,G33,G34,G35,G36,G37}).

After the work~\cite{Demozzi}, all these attempts in constructing models of
inflationary magnetogenesis have been believed to be untrustful because of the
so-called ``strong coupling problem'' and ``backreaction problem,'' and efforts
have been made in constructing successful scenarios in which both problems are
avoided~\cite{Ferreira,Membiela,Subramanian1,Caprini,Subramanian2,Tasinato,Campanelli2,Sasaki,Guo-Qian}.

Recently enough~\cite{Campanelli1}, however, we have stressed the fact that
there is a flaw in the arguments of~\cite{Demozzi} and in the
proposed nonstandard magnetogenesis mechanisms.
There, in fact, it is assumed that, after reheating, postinflationary electric currents froze inflation-produced
superhorizon magnetic fields. But this implies a violation of
causality since, as first pointed out in~\cite{Barrow-Tsagas},
postinflationary currents are generated by microphysical processes during reheating and, then,
are vanishing on superhorizon scales.
Fixing the ``causality flaw'' in the Ratra model by studying
the creation of photons out from the vacuum {\it via} the ``Parker mechanism''~\cite{Birrell-Davies,Parker-Toms},
we have found that the Ratra model is a viable, and indeed successful, scenario
of inflationary magnetogenesis.

The implications for cosmic magnetogenesis of having vanishing superhorizon electric currents
have been investigated by Tsagas~\cite{Tsagas1}. He analyzed the case of free Maxwell theory
in flat, marginally open, and marginally closed universes,
and a particular case of nonstandard electromagnetism, the one where magnetic fields during
inflation evolve as a power of the expansion parameter, $B \propto a^{-m}$ with $0 \leq m < 2$ (the latter case has been further developed in~\cite{Tsagas2}). The result is that, in general, superhorizon magnetic fields do not evolve adiabatically after reheating.

The aim of this paper is threefold. First, we will show that
quantum magnetic fluctuations in Maxwell theory evolve adiabatically
in a spatially flat universe once the Bunch-Davies vacuum is chosen
to be the physical vacuum state.
%in contrast with one of the results of~\cite{Tsagas1}.
Second, we will generalize the results in~\cite{Tsagas2} by studying the postinflationary evolution of superhorizon magnetic fields in a general nonconformal-invariant
electromagnetic theory.
Third, we will show that inflation-produced, superhorizon magnetic fields are superadiabatically
amplified after inflation in the Ratra model, thus supporting our recent claim that the Ratra model evades both the backreaction and strong-coupling problems.

\section{II. Free Maxwell Theory}

In this section, we analyze the evolution of magnetic fields in free Maxwell theory.
We use two equivalent approaches, the ``photon wave function''
and the ``magnetic flux'' approaches. These will also be used in the next section when we study the
case of nonconformal-invariant theories. While the first approach is, in our opinion, more direct,
the second one is useful when comparing our results with those of~\cite{Tsagas1,Tsagas2}.

\subsection{IIa. Photon wave function approach}

Let us consider the free Maxwell theory described by
the Lagrangian ${\mathcal L}_{\rm em}  = -\frac{1}{4} F_{\mu \nu} F^{\mu \nu}$,
where $F_{\mu \nu} = \partial_\mu A_\nu - \partial_\nu A_\mu$, and $A_\mu$ is the
photon field.
We restrict our analysis to the case of a spatially flat, Friedmann-Robertson-Walker universe,
described by the line element
\begin{equation}
\label{metric}
ds^2 = a^2(d\eta^2 - d \x^2),
\end{equation}
where $\eta$ is the conformal time and $a(\eta)$ is the expansion parameter,
which we normalize to unity at the present time $\eta_0$.
Working in the Coulomb gauge, $A_0 = \partial_i A_i = 0$, we quantize the
electromagnetic field by expanding it, in the Fock space, as
\begin{equation}
\label{F1}
{\A}(\eta,\x) = \sum_{\alpha=1,2} \int \!\! \frac{d^3k}{(2\pi)^3\sqrt{2k}} \, a_{\kk,\alpha} \,
A_{k,\alpha}(\eta) \, \ee_{\kk,\alpha} \, e^{i\kk \x} + \mbox{H.c.} ,
\end{equation}
where $A_\mu = (0,\A)$, $\kk$ is the comoving wave number, $k = |\kk|$, and $\ee_{\kk,\lambda}$ are the standard circular polarization vectors. The annihilation and creation operators $a_{\kk,\alpha}$ and $a_{\kk,\alpha}^{\dag}$
satisfy the usual commutation relations
$[a_{\kk,\alpha}, a_{\kk',\alpha'}^{\dag}] = (2\pi)^3 \delta_{\alpha \alpha'} \delta(\kk-\kk')$,
all the other commutators being null.
In order to get the usual commutation relations for the electromagnetic field and its canonical
conjugate momentum, one must impose the Wronskian condition
\begin{equation}
\label{Wronskian}
A_{k,\alpha} A'^{*}_{k,\alpha} - A^*_{k,\alpha} A'_{k,\alpha} = 2ik
\end{equation}
on the wave functions of the two photon polarization states $A_{k,\alpha}$.
(Hereafter, a prime indicates a differentiation with respect to the conformal time.)
Finally, the vacuum state $|0\rangle$ is defined by $a_{\kk,\lambda} |0\rangle = 0$ for all
$\kk$ and $\alpha$ and normalized as $\langle 0|0\rangle = 1$.
The photon wave functions satisfy the usual free harmonic oscillator equation,
\begin{equation}
\label{oscillator}
A_{k,\alpha}'' + k^2 A_{k,\alpha} = 0,
\end{equation}
whose solution is
\begin{equation}
\label{F2}
A_{k,\alpha}(\eta) = c_{1,\alpha}(k) \, e^{-ik\eta} + c_{2,\alpha}(k) \, e^{ik\eta}.
\end{equation}
Here, $c_{1,\alpha}(k)$ and $c_{2,\alpha}(k)$ are integration constants which are fixed
by the choice of the vacuum. The Wronskian condition implies that
\begin{equation}
\label{last}
|c_{1,\alpha}(k)|^2 - |c_{2,\alpha}(k)|^2 = 1.
\end{equation}
The physical vacuum is the so-called Bunch-Davies vacuum~\cite{Birrell-Davies} defined by
\footnote{If the vacuum state were different from the Bunch-Davies vacuum,
the evolution of quantum electromagnetic fluctuations could be very different from that
analyzed here~\cite{Green}. However, there are many arguments showing that
other kinds of allowed vacua, such as the $\alpha$-vacua~\cite{Chernikov},
are unphysical states (see, e.g.,~\cite{Banks,Susskind}).}
\begin{equation}
\label{F3}
c_{1,\alpha}(k) = 1 \, , \;\;\; c_{2,\alpha}(k) = 0.
\end{equation}
Accordingly, the photon wave functions, in free Maxwell theory,
are the usual plane waves
\begin{equation}
\label{plane-wave}
A_{k,\alpha}(\eta) = e^{-ik\eta}.
\end{equation}
Let us now introduce the magnetic field as usual as $a^2 \B = \nabla \times \A$.
The observable quantity is the vacuum expectation value (VEV) of the squared magnetic field
operator. Using Eq.~(\ref{F1}), we find
\begin{equation}
\label{spectrum}
\langle 0 |\B^2(\eta,\x)| 0 \rangle = \int_0^{\infty} \! \frac{dk}{k} \, B_k^2(\eta),
\end{equation}
where
\begin{equation}
\label{F4bis}
B_k^2(\eta) = \frac{k^4}{4\pi^2 a^4} \sum_{\alpha=1,2} |A_{k,\alpha}(\eta)|^2
\end{equation}
is the so-called magnetic power spectrum, and $B_k(\eta)$ the magnetic field on the scale $1/k$.
Inserting Eq.~(\ref{F2}) in Eq.~(\ref{F4bis}), and taking into account Eq.~(\ref{F3}), we find
\begin{equation}
\label{F5}
a^2 B_k(\eta) = \frac{k^2}{\sqrt{2}\pi} \, ,
\end{equation}
which shows that, in free Maxwell theory, magnetic fields evolve adiabatically
(i.e., proportionally to $a^{-2}$) for all times.
%, in contrast with the result of~\cite{Tsagas1}.

The introduction of the conductivity does not change this result. In fact,
its only effect, due to his huge value in the early universe, is to force
subhorizon magnetic fields to evolve adiabatically after reheating.
\footnote{Hereafter, we neglect possible effects of magnetohydrodynamic turbulence
that could be triggered by the electroweak and/or quark-hadron cosmological phase transitions,
and that could affect the evolution properties of inflation-produced magnetic
fields on subhorizon scales~\cite{MHD1,MHD2,MHD3,MHD4,MHD5,MHD6,MHD7,MHD8}.
For the cosmological relevant case of scaling-invariant fields (see Secs.~IVb and VIb), however, turbulence effects are suppressed in such a way that magnetic fields
stay almost unchanged on scales of cosmological interest~\cite{MHDInf1,MHDInf2}.}
Accordingly, magnetic fields that evolve adiabatically before reheating
will keep evolving adiabatically till today.

\subsection{IIb. Magnetic flux approach}

The ``magnetic flux'' is defined by $\BB(\eta,\x) = a^2 \B(\eta,\x)$.
In Fock space, we have
\begin{equation}
\label{T5tris}
{\BB}(\eta,\x) = \sum_{\alpha=1,2} \int \!\! \frac{d^3k}{(2\pi)^3\sqrt{2k}} \, a_{\kk,\alpha} \,
\BB_{\kk,\alpha}(\eta) \, e^{i\kk \x} + \mbox{H.c.} ,
\end{equation}
where we have defined
\begin{equation}
\label{Summing}
\BB_{\kk,\alpha} = k (-1)^{\alpha + 1} A_{k,\alpha}(\eta) \ee_{\kk,\alpha}.
\end{equation}
Summing up the two photon polarization states, $\sum_{\alpha=1,2} \BB_{\kk,\alpha}$,
we get the Fourier transform of the magnetic flux. However, it is more useful to introduce
the quantity
\footnote{Apart from inessential numerical factors, the components of $\B_{\kk}$ and $\BB_{\kk}$ coincide, respectively, with the quantities $B_{(n)}$ and $\mathcal{B}_{(n)}$ defined in~\cite{Tsagas1,Tsagas2} for
the case of a spatially flat universe.}
\begin{equation}
\label{T11}
a^2 \B_{\kk} = \BB_{\kk} = \frac{k}{2\pi} \sum_{\alpha=1,2} \BB_{\kk,\alpha}
\end{equation}
(which we will still call Fourier transform of the magnetic flux for the sake of convenience),
for two reasons. First, it has the dimension of a magnetic field and, second,
because
\begin{equation}
\label{T5bis}
|\B_{\kk}(\eta)| = B_k(\eta),
\end{equation}
namely, the modulus of the Fourier-transformed magnetic field is equal to the magnetic field on the scale $1/k$.

The Fourier transform of the magnetic flux satisfies,
in the hypothesis of null conductivity, the field equation
\begin{equation}
\label{T6}
\BB''_{\kk} + k^2 \BB_{\kk} = 0,
\end{equation}
whose solution is
\begin{equation}
\label{T7}
\BB_{\kk} = a^2 \B_{\kk} = \C_1 \cos(k\eta) + \C_2 \sin(k\eta),
\end{equation}
where $\C_1(\kk)$ and $\C_2(\kk)$ are complex constant  vectors of integration.
Inserting Eq.~(\ref{F2}) in Eq.~(\ref{Summing}), and
comparing the resulting expression with Eq.~(\ref{T7}), we get
\begin{eqnarray}
\label{T12}
\!\!\!\!\!\!\!\!\!\! && \C_1 = \frac{k^2}{2\pi} \sum_\alpha (-1)^{\alpha + 1} [c_{1,\alpha}(k) + c_{2,\alpha}(k)] \ee_{\kk,\alpha}, \\
\label{T13}
\!\!\!\!\!\!\!\!\!\! && \C_2 = -i\frac{k^2}{2\pi} \sum_\alpha (-1)^{\alpha + 1} [c_{1,\alpha}(k) - c_{2,\alpha}(k)] \ee_{\kk,\alpha}.
\end{eqnarray}
The Wronskian condition~(\ref{Wronskian}) on the photon wave function implies that
\begin{equation}
\label{Wronskian2}
\C_1 \cdot \C_2^* - \C_1^* \cdot \C_2 = \frac{ik^4}{\pi} \, .
\end{equation}
Equations~(\ref{T12})-(\ref{T13}) show that the integration constants $\C_1$ and $\C_2$ are fixed
by the choice of the vacuum.
The (physical) Bunch-Davies vacuum is defined by Eq.~(\ref{F3}), and then it corresponds to take
\begin{equation}
\label{T14}
\C_1 = \frac{k^2}{2\pi} \left(\ee_{\kk,1} - \ee_{\kk,2} \right), \;\;\; \C_2 = -i \C_1.
\end{equation}
Inserting the above relations in Eq.~(\ref{T7}), we get
\begin{equation}
\label{T14bis}
a^2 \B_{\kk}(\eta) = \C_1 e^{-ik\eta},
\end{equation}
from which it follows that $|a^2 \B_{\kk}(\eta)| = k^2/\sqrt{2} \pi$,
in agreement with Eq.~(\ref{F5}). %This result differs from that obtained in~\cite{Tsagas1}.

To compare our results with those in~\cite{Tsagas1}, let us rewrite Eq.~(\ref{T7}) in the long wavelength limit (namely, for superhorizon modes, $|k\eta| \ll 1$),
\begin{equation}
\label{T8}
a^2 \B_{\kk} \simeq \C_1 + \C_2 k\eta.
\end{equation}
From the above equation, it follows that the constant $\C_1$ and $\C_2$ can be expressed as a
function of $\B_{\kk}$ and its first derivative calculated at a reference time $\eta_*$:
\begin{eqnarray}
\label{T9}
\!\!\!\!\!\!\!\!\!\! && \C_1 \simeq -[(2\eta \mathcal{H} -1) \B_{\kk} + \eta \B_{\kk}'] \, a^2|_{\eta = \eta_*}, \\ %+\mathcal{O}(k\eta_*), \\
\label{T10}
\!\!\!\!\!\!\!\!\!\! && \C_2 \simeq [2\eta \mathcal{H} \B_{\kk} + \eta \B_{\kk}'] \, \frac{a^2}{k\eta}|_{\eta = \eta_*},
\end{eqnarray}
where $\mathcal{H} = a'/a$.
Now the author of~\cite{Tsagas1} argues that, since $\C_2$ is inversely proportional to $k\eta_*$,
then $|\C_2| \gg |\C_1|$, unless the quantity $2\eta \mathcal{H} \B_{\kk} + \eta \B_{\kk}'$ is null.
Consequently, the dominant term in  Eq.~(\ref{T8}) would be the nonadiabatic one (namely that proportional to $\C_2$), and this would open the possibility to have a superadiabatic evolution of magnetic fields in the free Maxwell theory. However, this does not happen in a theory based on the Bunch-Davies vacuum
since the quantity $2\eta \mathcal{H} \B_{\kk} + \eta \B_{\kk}'$
is, in this case, null at the lowest order in $|k\eta| \ll 1$.
In fact, from Eq.~(\ref{T14bis}), it follows that
\begin{equation}
\label{T15}
2\eta \mathcal{H} \B_{\kk} + \eta \B_{\kk}' = -ik\eta \B_{\kk}.
\end{equation}
When inserted in Eqs.~(\ref{T9})-(\ref{T10}), the above equation gives
$\C_1 \simeq i \C_2 \simeq a^2 \B_{\kk}|_{\eta = \eta_*}$,
which shows that $|\C_1|$ and $|\C_2|$ are constants of the same magnitude
and, in turn, that the magnetic field evolves adiabatically on superhorizon scales.
Finally, for the sake of completeness, we observe that the exact result is
\begin{equation}
\label{J1}
\C_1 = i\C_2 = e^{ik\eta} a^2 \B_{\kk}|_{\eta = \eta_*}.
\end{equation}

\section{III. Nonconformal-invariant theories}

Let us now consider the case where electromagnetic fields during
inflation are described by a (nonstandard) nonconformal-invariant electromagnetic Lagrangian.
After inflation, instead, we assume that such a Lagrangian smoothly reduces to the Maxwell
Lagrangian in order to recover standard electromagnetism and not to spoil, thus,
the predictions of the standard cosmological model.

\subsection{IIIa. Photon wave function approach}

Since photons after inflation are described by the standard free electromagnetism,
they evolve according to Eq.~(\ref{F2}), with $c_{1,\alpha}(k)$ and $c_{2,\alpha}(k)$
being complex functions of $k$ that are fixed by the properties of the electromagnetic field
at the end of inflation. Let us assume, for the sake of simplicity, that the dynamics of the
electromagnetic field during inflation is parity conserving, so that $c_{1,\alpha}(k)$ and $c_{2,\alpha}(k)$ do not depend on the photon helicity index $\alpha$:
\begin{equation}
\label{BD}
c_{1,\alpha}(k) = \alpha_k, \;\;\; c_{2,\alpha}(k) = \beta_k.
\end{equation}
(The parity-violating case, which is associated with the production of magnetic helicity,
goes along the same lines as below.)
The coefficients $\alpha_k$ and $\beta_k$ are known as the Bogolubov coefficient, and
the Wronskian condition implies the Bogolubov relation
\begin{equation}
\label{Wronskian3}
|\alpha_k|^2 - |\beta_k|^2 = 1.
\end{equation}
The square modulus of the coefficient $\beta_k$ gives the number (density)
of the produced photons during inflation~\cite{Campanelli1},
\begin{equation}
\label{BDbis}
n_\kk = |\beta_k|^2,
\end{equation}
and it is zero for the case of conformal-invariant theories~\cite{Birrell-Davies},
such as the free Maxwell theory in a Friedmann-Robertson-Walker spacetime.
Inserting Eq.~(\ref{F2}) in Eq.~(\ref{F4bis}), and taking into account Eqs.~(\ref{BD}), (\ref{Wronskian3})
and (\ref{BDbis}), we find
\begin{equation}
\label{Io1}
a^2 B_k(\eta) = \! \frac{k^2}{\sqrt{2}\pi} \sqrt{1 \! + \! 2n_\kk \! + \! 2 \sqrt{n_\kk(n_\kk \! + \! 1)} \cos(\Omega_k \! - \! 2k\eta)},
\end{equation}
where
\begin{equation}
\label{y2}
\Omega_k = \mbox{Arg} (\alpha_k \beta_k^*) \, \in \; ]-\pi,\pi]
\end{equation}
is the phase difference between the two Bogolubov coefficients.
Equation~(\ref{Io1}) is in agreement with the result of~\cite{Campanelli1}.

If $n_\kk = 0$ (conformal-invariant theories), then $a^2 B_k(\eta) = k^2/\sqrt{2}\pi$,
in agreement with the result of Sec.~II [see Eq.~(\ref{F5})].

In general, in order to explain the large-scale magnetic fields we observe today in galaxies and clusters of galaxies,
we must have $n_\kk \gg 1$ on scales $\lambda = 1/|\kk|$ of astrophysical interest for cosmic magnetic fields (see Sec.~V).
Since we are interested in the evolution of superhorizon modes,
let us expand Eq.~(\ref{Io1}) in the limit $n_\kk \gg 1$ and $|k\eta| \ll 1$. Independently on the order of the expansion,
and to the lowest order, we find
\begin{equation}
\label{Io2}
a^2 B_k(\eta) \simeq \frac{k^2}{\pi} \sqrt{(1 + \cos\Omega_k) n_\kk} \, .
\end{equation}
Therefore, a necessary condition for having a nonadiabatic evolution of
superhorizon magnetic fields after inflation is that $\Omega_k \rightarrow \pm \pi$
for $n_\kk \gg 1$. Let us define, for later convenience, the two cases
\begin{eqnarray}
\label{Q1}
&& \mbox{case A1} \!: \;\; |\Omega_k \mp \pi| = \mathcal{O}(1) \;\; \mbox{and} \;\; n_\kk^{-1} \ll |k\eta| \ll 1, \nonumber \\
&& \mbox{case A2} \!: \;\; |\Omega_k \mp \pi| = \mathcal{O}(1) \;\; \mbox{and} \;\; |k\eta| \ll n_\kk^{-1} \ll 1. \nonumber \\
\end{eqnarray}
When cases A1 and A2 are realized, the evolution is then adiabatic.

Let us now consider superhorizon modes such that $\Omega_k \rightarrow \pm \pi$ for $n_\kk \gg 1$.
We have six cases: $n_\kk^{-1} \ll |\Omega_k \mp \pi| \ll |k\eta|$ and cyclic permutations.
After expanding Eq.~(\ref{Io1}), we have, to the lowest order,
\begin{equation}
\label{Io3}
a^2 B_k(\eta) \simeq \frac{k^2}{\sqrt{2}\pi} \times
\left\{ \begin{array}{lll}
  2\sqrt{n_\kk} \, |k \eta|,           & \mbox{B1, B2},
  \\
  |\Omega_k \mp \pi| \sqrt{n_\kk} \, , & \mbox{C1, C2},
  \\
  \frac12 \sqrt{1/n_\kk},              & \mbox{D1, D2},
  \end{array}
  \right.
\end{equation}
where the six cases, B1, B2, C1, C2, D1, and D2, correspond to
\begin{eqnarray}
\label{Q2}
&& \mbox{case B1} \!: \;\; n_\kk^{-1} \ll |\Omega_k \mp \pi| \ll |k\eta| \ll 1, \nonumber \\
&& \mbox{case B2} \!: \;\; |\Omega_k \mp \pi| \ll n_\kk^{-1} \ll |k\eta| \ll 1, \nonumber \\
&& \mbox{case C1} \!: \;\; n_\kk^{-1} \ll |k\eta| \ll |\Omega_k \mp \pi| \ll 1, \nonumber \\
&& \mbox{case C2} \!: \;\; |k\eta| \ll n_\kk^{-1} \ll |\Omega_k \mp \pi| \ll 1, \nonumber \\
&& \mbox{case D1} \!: \;\; |\Omega_k \mp \pi| \ll |k\eta| \ll n_\kk^{-1} \ll 1, \nonumber \\
&& \mbox{case D2} \!: \;\; |k\eta| \ll |\Omega_k \mp \pi| \ll n_\kk^{-1} \ll 1, \nonumber \\
\end{eqnarray}
respectively. Looking at Eq.~(\ref{Io3}), we conclude that the evolution of superhorizon magnetic
fields is superadiabatic, $a^2B_k(\eta) \propto \eta$, only in the cases B1 and B2,
while it is adiabatic in the remaining cases.

Let us now follow the evolution of superhorizon magnetic fields from the end of inflation,
at $\eta = \eta_e < 0$, until today. Let us observe that, although the function $a^2 B_k(\eta)$
in Eq.~(\ref{Io1}) in not an even function of the conformal time, its asymptotic expansions
in Eqs.~(\ref{Io2}) and (\ref{Io3}) are. This allows us to simplify the problem and
to consider such an evolution from the positive time $|\eta_e|$ till the present time $\eta_0$.

If a given magnetic mode with wave number $k$ starts his
evolution in the case A1, namely if $|\Omega_k \mp \pi| = \mathcal{O}(1)$ and $n_\kk^{-1} \ll |k\eta_e| \ll 1$,
then as the time passes and $\eta$ grows, it will remain in the case A1 until it reenters the horizon at
the time $\eta_\downarrow$ defined by $k\eta_\downarrow \simeq 1$. After that, its evolution is still adiabatic because of the high conductivity which freezes any subhorizon magnetic field into the primeval plasma.
The evolution, then, is adiabatic for all times.

If the magnetic mode starts in the case A2, its evolution will always be adiabatic,
although after a time of order $\eta_{\tiny \mbox{A2} \rightarrow \mbox{A1}} \sim 1/k n_\kk$,
it will move from the case A2 to the case A1.

If the magnetic mode starts in either the case B1 or B2,
its evolution will be superadiabatic up to the time $\eta_\downarrow$ of its reentering the horizon,
and from that time on it will evolve adiabatically.

If the magnetic mode starts in the case C1, it will evolve adiabatically up to the time of order
$\eta_{\tiny \mbox{C1} \rightarrow \mbox{B1}} \simeq |\Omega_k \mp \pi|/2k$, after which it will move
in the case B1, and then will evolve superadiabatically up to its reentering the horizon.

%************************************   Figure 1   *******************************************%

\begin{figure*}[t!]
\begin{center}
\includegraphics[clip,width=0.48\textwidth]{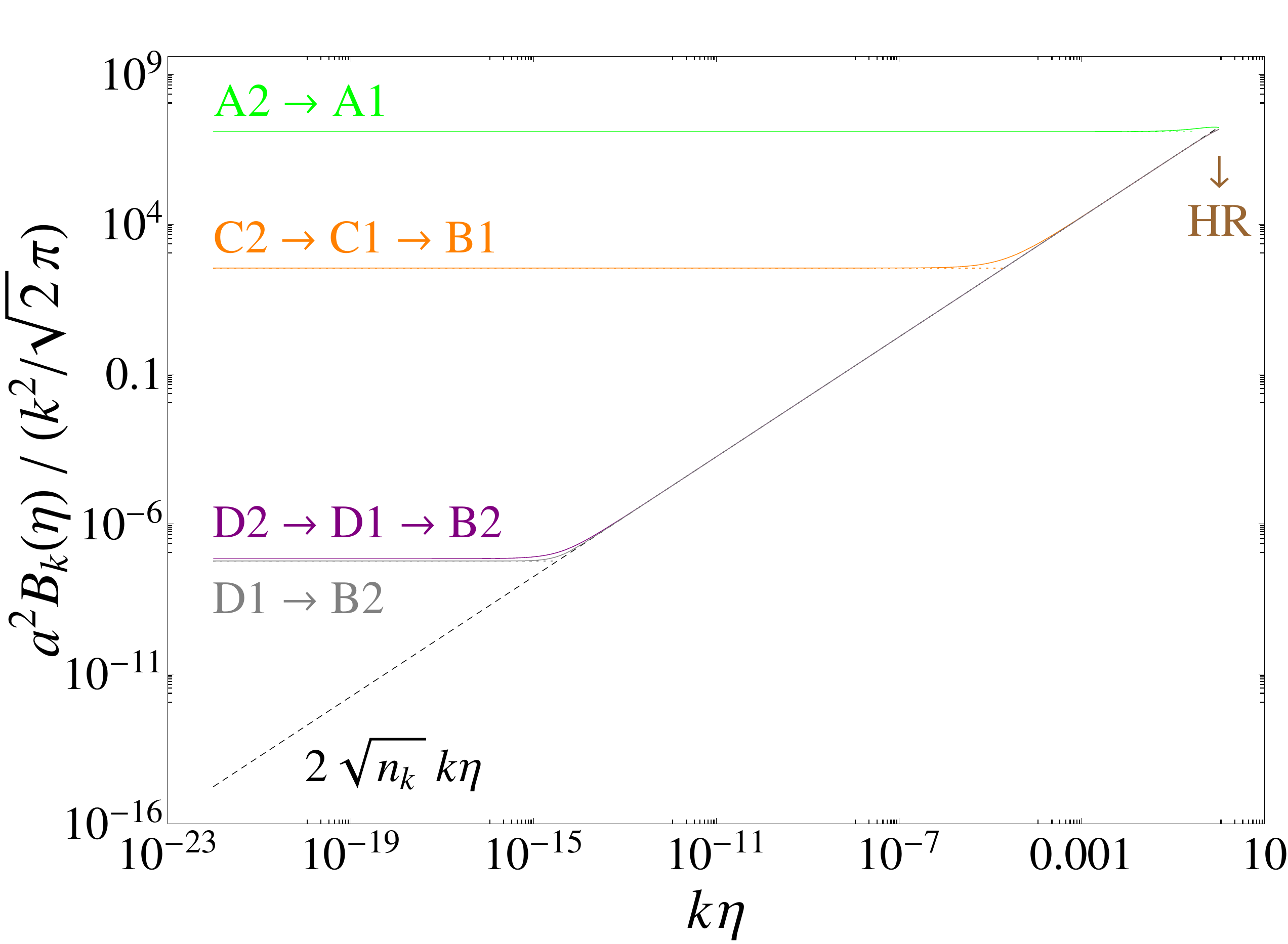}
\hspace{0.5cm}
\includegraphics[clip,width=0.48\textwidth]{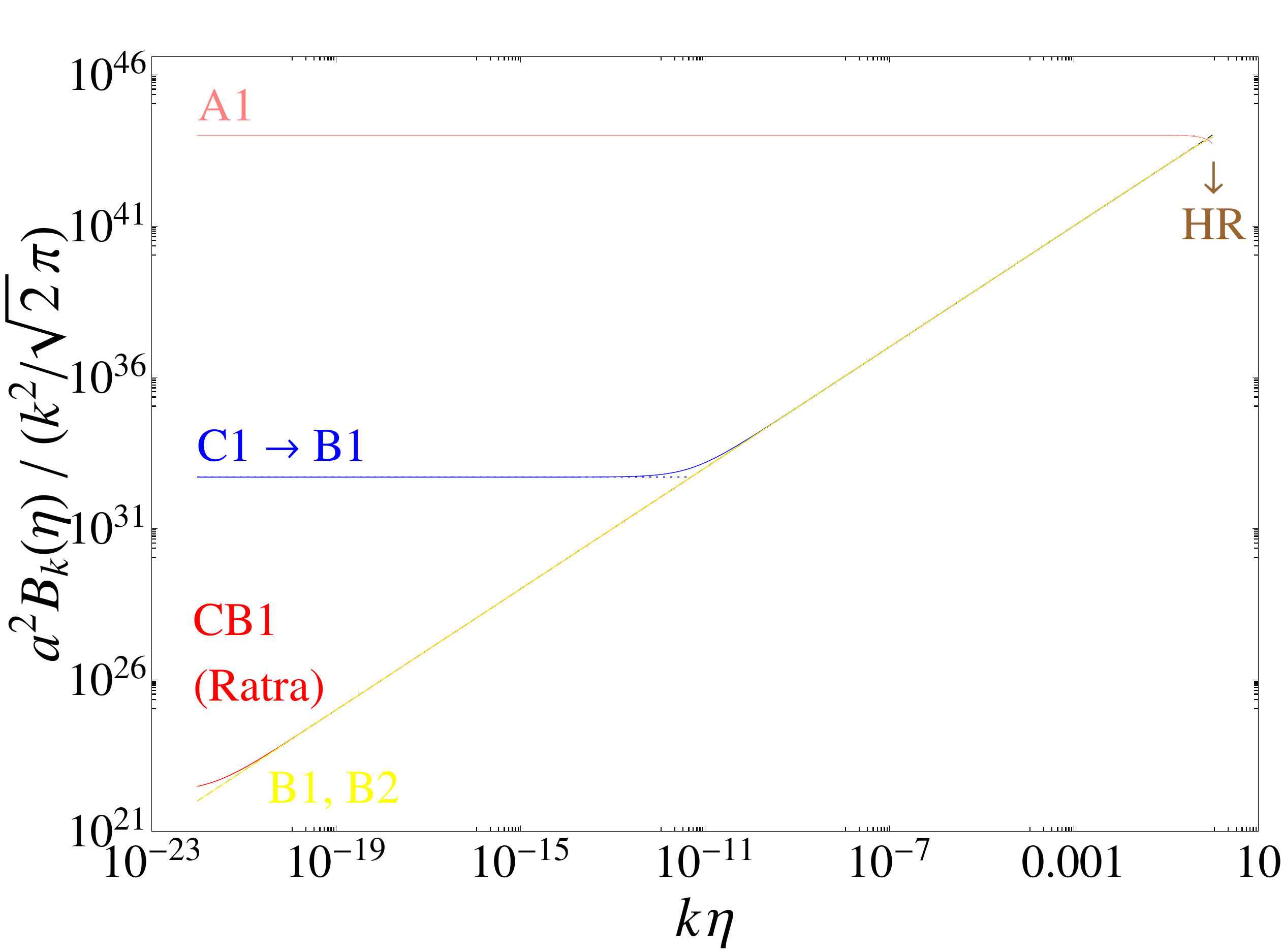}
\caption{The magnetic flux spectrum, $a^2 B_k(\eta)$, in a nonconformal-invariant theory normalized to its expression in the case of free Maxwell theory, $k^2/\sqrt{2}\pi$, as a function of $k\eta$
for different initial conditions at the end of inflation ($\eta = |\eta_e|$). For a given magnetic mode $k$, such conditions are determined by the values of $k|\eta_e|$, $n_\kk$, and $\Omega_k$, where
$n_\kk$ is the number density of photons created during inflation and $\Omega_k$ is the
phase difference between the two Bogolubov coefficients which define the state of the magnetic field
at the end of inflation.
Continuous lines refer to the exact expression of the magnetic flux [see Eq.~(\ref{Io1})],
while dashed lines to its asymptotic expansions [see Eqs.~(\ref{Io2}) and (\ref{Io3})].
The cases A1, A2, B1, B2, C1, C2, D1, D2 are defined in Eqs.~(\ref{Q1}) and (\ref{Q2}), and
``HR'' stands for ``horizon reentering''.
{\it Left panel.} $k|\eta_e| = 10^{-22}$ and $n_\kk = (2/5)^2|k\eta_e|^{-2/3}$. From top to bottom,
$\Omega_k = \pi/2$ ($\mbox{A2} \rightarrow \mbox{A1}$),
$\Omega_k = \pi - |k\eta_e|^{1/5}$ ($\mbox{C2} \rightarrow \mbox{C1} \rightarrow \mbox{B1}$),
$\Omega_k = \pi - 2|k\eta_e|^{2/3}$ ($\mbox{D2} \rightarrow \mbox{D1} \rightarrow \mbox{B2}$), and
$\Omega_k = \pi - |k\eta_e|^{4}$ ($\mbox{D1} \rightarrow \mbox{B2}$).
{\it Right panel.} $k|\eta_e| = 10^{-22}$ and $n_\kk = (1/4)|k\eta_e|^{-4}$. From top to bottom,
$\Omega_k = 0$ (A1),
$\Omega_k = \pi - |k\eta_e|^{1/2}$ ($\mbox{C1} \rightarrow \mbox{B1}$),
$\Omega_k$ given in Eq.~(\ref{RR9}) (Ratra model discussed in Sec.~IVb),
$\Omega_k = \pi - |k\eta_e|^{3}$ (B1), and
$\Omega_k = \pi - |k\eta_e|^{5}$ (B2).}
\end{center}
\end{figure*}

%*********************************************************************************************%

If a mode starts in the case C2, it will first evolve adiabatically, then it will move from the case C2 to the case C1
at the time $\eta_{\tiny \mbox{C2} \rightarrow \mbox{C1}} \sim 1/k n_\kk$, then will continue its adiabatic evolution
up to the time $\eta_{\tiny \mbox{C1} \rightarrow \mbox{B1}}$, after which it will evolve superadiabatically
according to the case B1 until it reenters the horizon.

If the magnetic mode starts in the case D1, it will evolve adiabatically up to the time of order
$\eta_{\tiny \mbox{D1} \rightarrow \mbox{B2}} \simeq 1/4kn_\kk$, after which it will move
in the case B2, and then will evolve superadiabatically up to its reentering the horizon.

If a mode starts in the case D2, it will first evolve adiabatically, then it will move from the case D2 to the case D1
at the time $\eta_{\tiny \mbox{D2} \rightarrow \mbox{D1}} \sim |\Omega_k \mp \pi|/k$, then will continue its adiabatic evolution
up to the time $\eta_{\tiny \mbox{D1} \rightarrow \mbox{B2}}$, after which it will evolve superadiabatically
according to the case B2 until it reenters the horizon.

Schematically, the evolution of a given magnetic mode is as follows:
\begin{eqnarray}
\label{D1}
&& \mbox{A1} \rightarrow \mbox{HR} \nonumber \\
&& \mbox{A2} \rightarrow \mbox{A1} \rightarrow \mbox{HR} \nonumber \\
&& \mbox{B1} \rightarrow \mbox{HR} \nonumber \\
&& \mbox{B2} \rightarrow \mbox{HR} \nonumber \\
&& \mbox{C1} \rightarrow \mbox{B1} \rightarrow \mbox{HR} \nonumber \\
&& \mbox{C2} \rightarrow \mbox{C1} \rightarrow \mbox{B1} \rightarrow \mbox{HR} \nonumber \\
&& \mbox{D1} \rightarrow \mbox{B2} \rightarrow \mbox{HR} \nonumber \\
&& \mbox{D2} \rightarrow \mbox{D1} \rightarrow \mbox{B2} \rightarrow \mbox{HR} \nonumber
\end{eqnarray}
where HR stands for horizon reentering.
\footnote{The full evolution of the inflation-produced magnetic field from the
end of inflation at $\eta_e < 0$ up to its reentering the horizon at $\eta_\downarrow > 0$
can be schematically described as follows:
\begin{eqnarray}
\label{D1}
&& \mbox{A1} \rightarrow \mbox{A2} \rightarrow \mbox{A1} \rightarrow \mbox{HR} \nonumber \\
&& \mbox{A2} \rightarrow \mbox{A1} \rightarrow \mbox{HR} \nonumber \\
&& \mbox{B1} \rightarrow \mbox{C1} \rightarrow \mbox{C2} \rightarrow \mbox{C1} \rightarrow \mbox{B1} \rightarrow \mbox{HR} \nonumber \\
&& \mbox{B2} \rightarrow \mbox{D1} \rightarrow \mbox{D2} \rightarrow \mbox{D1} \rightarrow \mbox{B2} \rightarrow \mbox{HR} \nonumber \\
&& \mbox{C1} \rightarrow \mbox{C2} \rightarrow \mbox{C1} \rightarrow \mbox{B1} \rightarrow \mbox{HR} \nonumber \\
&& \mbox{C2} \rightarrow \mbox{C1} \rightarrow \mbox{B1} \rightarrow \mbox{HR} \nonumber \\
&& \mbox{D1} \rightarrow \mbox{D2} \rightarrow \mbox{D1} \rightarrow \mbox{B2} \rightarrow \mbox{HR} \nonumber \\
&& \mbox{D2} \rightarrow \mbox{D1} \rightarrow \mbox{B2} \rightarrow \mbox{HR} \nonumber
\end{eqnarray}
where the cases A1, A2, B1, B2, C1, C2, D1, D2 are the same as in Eqs.~(\ref{Q1}) and (\ref{Q2}).}

It is interesting to observe that the actual magnetic field, $B_k(\eta_0)$,
which coincides with the magnetic flux at the time of horizon reentering,
$a^2 B_k(\eta)|_{\eta = \eta_\downarrow}$, is independent of the details of its
evolution when outside the horizon. Indeed, from Eq.~(\ref{Io1}), we find, in the limit $n_\kk \gg 1$,
\begin{equation}
\label{D4}
B_k(\eta_0) \simeq  \zeta_k k^2 \sqrt{n_\kk} \, ,
\end{equation}
where $\zeta_k = \sqrt{1 + \cos(\Omega_k -2k\eta_\downarrow)}/\pi$.
Since $k\eta_\downarrow \simeq 1$ by definition, we see that,
excluding a region very near to $\Omega_k \simeq 2- \pi$ (where $\zeta_k$
is vanishing), $\zeta_k$ is an order-one function $\Omega_k$.
Therefore, the actual magnetic field spectrum is essentially determined by
the number of photons with wave number $k$ that have been created during inflation.

In Fig.~1, we show the magnetic flux spectrum $a^2 B_k(\eta)$ normalized to its expression in the
case of pure Maxwell theory, $k^2/\sqrt{2}\pi$ [see Eq.~(\ref{F5})], as a function of $k\eta$
(with $\eta > 0$) for different initial conditions at the end of inflation at $\eta = |\eta_e|$.
Continuous lines refer to the exact expression of the magnetic flux given in Eq.~(\ref{Io1}),
while dashed lines to its asymptotic expansions given in Eqs.~(\ref{Io2}) and (\ref{Io3}).
In both panels, we have taken $k|\eta_e| = 10^{-22}$.
In the left panel, $n_\kk = (2/5)^2|k\eta_e|^{-2/3}$ and,
from top to bottom,
$\Omega_k = \pi/2$ ($\mbox{A2} \rightarrow \mbox{A1}$),
$\Omega_k = \pi - |k\eta_e|^{1/5}$ ($\mbox{C2} \rightarrow \mbox{C1} \rightarrow \mbox{B1}$),
$\Omega_k = \pi - 2|k\eta_e|^{2/3}$ ($\mbox{D2} \rightarrow \mbox{D1} \rightarrow \mbox{B2}$), and
$\Omega_k = \pi - |k\eta_e|^{4}$ ($\mbox{D1} \rightarrow \mbox{B2}$).
In the right panel, $n_\kk = (1/4)|k\eta_e|^{-4}$ and
from top to bottom,
$\Omega_k = 0$ (A1),
$\Omega_k = \pi - |k\eta_e|^{1/2}$ ($\mbox{C1} \rightarrow \mbox{B1}$),
$\Omega_k$ given in Eq.~(\ref{RR9}) (scaling-invariant case in the Ratra model discussed in Sec.~IVb),
$\Omega_k = \pi - |k\eta_e|^{3}$ (B1), and
$\Omega_k = \pi - |k\eta_e|^{5}$ (B2). %(the last two coincide graphically).

\subsection{IIIb. Magnetic flux approach}

Let us now investigate the evolution of superhorizon magnetic
modes using the magnetic flux approach. Taking into account Eq.~(\ref{BD}), the coefficients
$\C_1$ and $\C_2$ in Eqs.~(\ref{T12}) and (\ref{T13}) are
\begin{eqnarray}
\label{x1}
&&  \C_1 = \frac{k^2}{2\pi} \, (\alpha_k + \beta_k) \left(\ee_{\kk,1} - \ee_{\kk,2} \right), \\
\label{x2}
&& \C_2 = -i \frac{k^2}{2\pi} \, (\alpha_k - \beta_k) \left(\ee_{\kk,1} - \ee_{\kk,2} \right).
\end{eqnarray}
Inserting these expression in Eq.~(\ref{T7}), we find
\begin{equation}
\label{T14tris}
a^2 \B_{\kk}(\eta) = \C_1 \left( \frac{\alpha_k}{\alpha_k + \beta_k} \, e^{-ik\eta} +
\frac{\beta_k}{\alpha_k + \beta_k} \, e^{ik\eta}\right) \! ,
\end{equation}
from which it follows that
\begin{equation}
\label{x6}
2\eta \mathcal{H} \B_{\kk} + \eta \B_{\kk}' = -ik\eta \B_{\kk} \, g(n_\kk,\Omega_k,k\eta),
\end{equation}
where we have introduced the function
\begin{equation}
\label{x6bis}
g(n_\kk,\Omega_k,k\eta) =
\frac{e^{i(\Omega_k -2k\eta)} \sqrt{n_\kk + 1} - \sqrt{n_\kk}}{e^{i(\Omega_k -2k\eta)} \sqrt{n_\kk + 1} + \sqrt{n_\kk}}  \, .
\end{equation}
Inserting Eq.~(\ref{x6}) in Eqs.~(\ref{T9})-(\ref{T10}), we find
\begin{eqnarray}
\label{new1}
&& \C_1 \simeq (1+ik\eta g) a^2 \B_{\kk}|_{\eta = \eta_*}, \\
\label{new2}
&& \C_2 \simeq - i g a^2 \B_{\kk}|_{\eta = \eta_*}.
\end{eqnarray}
For the eight cases discussed above, we find, to the lowest order,
\begin{equation}
\label{x10}
g(n_\kk,\Omega_k,k\eta) \simeq
\left\{ \begin{array}{lll}
  i \tan(\Omega_k/2),     & \mbox{A1, A2}
  \\
  i/k \eta,               & \mbox{B1, B2},
  \\
  -2i/(\Omega_k \mp \pi), & \mbox{C1, C2},
  \\
  4n_\kk,                 & \mbox{D1, D2}.
  \end{array}
  \right.
\end{equation}
Inserting Eq.~(\ref{x10}) in Eqs.~(\ref{new1})-(\ref{new2}), we find
\footnote{In order to find the infinitesimal term that corresponds to ``0'' in the first equation of Eq.~(\ref{c2}),
we need to go to the next order in the expansion of the function $g$ in Eq.~(\ref{x10}). In this case,
we find $\C_1 \simeq -(|\Omega_k \mp \pi|/2k\eta_* \!) \, a^2 \B_{\kk}|_{\eta = \eta_*}$ for the case B1,
and $\C_1 \simeq (i/4 n_\kk k\eta_* \!) \, a^2 \B_{\kk}|_{\eta = \eta_*}$ for the case B2.
We arrive at the same results for $\C_1$ and $\C_2$ in Eqs.~(\ref{c1})-(\ref{c4}) if we first consider their exact expressions (see below), and then we expand them in terms of $n_\kk$, $\Omega_k$, and $k\eta_*$.}
\begin{eqnarray}
\label{c1}
\!\!\!\!\!\!\!\!\!\!\! && \C_1 \simeq \cot(\Omega_k/2) \, \C_2 \simeq a^2 \B_{\kk}|_{\eta = \eta_*}, ~~~~~~\,\,     \mbox{A1, A2},  \\
\label{c2}
\!\!\!\!\!\!\!\!\!\!\! && \C_1 \simeq 0, \;\;\;  \C_2 \simeq (k\eta_*)^{-1} a^2 \B_{\kk}|_{\eta = \eta_*}, ~~~~\,\, \mbox{B1, B2}, \\
\label{c3}
\!\!\!\!\!\!\!\!\!\!\! && \C_1 \simeq -[(\Omega_k \mp \pi)/2] \, \C_2 \simeq  a^2 \B_{\kk}|_{\eta = \eta_*},    ~\, \mbox{C1, C2}, \\
\label{c4}
\!\!\!\!\!\!\!\!\!\!\! && \C_1 \simeq (i/4n_\kk) \, \C_2 \simeq  a^2 \B_{\kk}|_{\eta = \eta_*},       ~~~~~~~~~\:\, \mbox{D1, D2}.
\end{eqnarray}
The exact expressions for $\C_1$ and $\C_2$ come from the expression of $a^2 \B_{\kk}$ in Eq.~(\ref{T14tris})
evaluated at $\eta = \eta_*$ and from Eq.~(\ref{x2}). They are:
\begin{eqnarray}
\label{J2}
&& \C_1 = \frac{a^2 \B_{\kk}|_{\eta = \eta_*}}{\cos(k\eta_*) - i g(n_\kk,\Omega_k,0) \sin(k\eta_*)} \, , \\
\label{J3}
&& \C_2 = -ig(n_\kk,\Omega_k,0) \C_1.
\end{eqnarray}
These expressions reduce, in the limit $n_\kk = 0$,
to those previously found for the free Maxwell theory [see Eq.~(\ref{J1})].
Inserting Eqs.~(\ref{c1})-(\ref{c4}) in Eq.~(\ref{T8}), we find
\begin{equation}
\label{BB1}
\;\, a^2 \B_{\kk}(\eta) \simeq \!
\left\{ \begin{array}{lll}
  \!\!a^2 \B_{\kk}|_{\eta = \eta_*}, & \!\mbox{A1, A2, C1, C2, D1, D2},
  \\
  \!\!a^2 \B_{\kk}|_{\eta = \eta_*} \left(\frac{\eta}{\eta_*}\right)\! ,  & \!\mbox{B1, B2}.
  \end{array}
  \right.
\end{equation}
Finally, observing that $|a^2 \B_{\kk}(\eta)| \simeq |\C_1|$ for the cases A1, A2, C1, C2, D1, and D2,
and $|a^2 \B_{\kk}(\eta)| \simeq |\C_2| |k\eta|$ for the cases B1, B2, we obtain, at the time $\eta = \eta_*$,
\begin{equation}
\label{BB2}
|a^2 \B_{\kk}|_{\eta = \eta_*} \simeq \frac{k^2}{\sqrt{2}\pi} \times
\left\{ \begin{array}{lll}
  \sqrt{2(1 + \cos\Omega_k) n_\kk} \, , & \mbox{A1, A2},
  \\
  2\sqrt{n_\kk} \, |k \eta_*|,          & \mbox{B1, B2},
  \\
  |\Omega_k \mp \pi| \sqrt{n_\kk} \, ,  & \mbox{C1, C2},
  \\
  \frac12 \sqrt{1/n_\kk},               & \mbox{D1, D2},
  \end{array}
  \right.
\end{equation}
where we used the following asymptotic expansions of $\C_1$ and $\C_2$ in Eqs.~(\ref{x1})-(\ref{x2}):
$|\C_1|^2 \simeq k^4 (1 + \cos\Omega_k) n_\kk/\pi^2$ for the cases A1, A2;
$|\C_2|^2 \simeq 2k^4 n_\kk/\pi^2$ for the cases B1, B2;
$|\C_1|^2 \simeq k^4 (\Omega_k \mp \pi)^2 n_\kk/2\pi^2$ for the cases C1, C2; and
$|\C_1|^2 \simeq k^4/8\pi^2n_\kk$ for the cases D1, D2.

Taking the modulus of $a^2 \B_{\kk}(\eta)$ in Eq.~(\ref{BB1}), and taking into account Eq.~(\ref{BB2}),
we arrive at the same results previously obtained using the photon wave function approach
[see Eqs.~(\ref{Io2}) and (\ref{Io3})].

\section{IV. Specific models}

We analyze now two specific models of inflationary magnetogenesis.
The first one is general enough and just assumes that
superhorizon magnetic fields evolve as a power of the conformal time during inflation,
while the second one is the well-known Ratra model.

\subsection{IVa. Power-law magnetic fields during inflation}

Following~\cite{Tsagas1,Tsagas2}, we rewrite Eqs.~(\ref{T8}), (\ref{T9}), and (\ref{T10}) as
\begin{eqnarray}
\label{BB3}
\B_{\kk}(\eta) \!\!& = &\!\! -[(2\eta \mathcal{H} -1) \B_{\kk} + \eta \B_{\kk}']_{\eta = \eta_e}
\left(\frac{a_e}{a}\right)^{\!2} \nonumber \\
\!\!& + &\!\! [2\eta \mathcal{H} \B_{\kk} + \eta \B_{\kk}']_{\eta = \eta_e}
\left(\frac{a_e}{a}\right)^{\!2} \! \left(\frac{\eta}{\eta_e}\right) \! ,
\end{eqnarray}
where $a_e = a(\eta_e)$. Let us specialize Eq.~(\ref{BB3}) to the case where inflation
is described by a pure de Sitter phase, or by a slow-roll phase characterized by a slow-roll parameter $\epsilon_1$, or by a power-law expansion
$a(t) \propto t^q$ with $q > 1$~\cite{Inflation}, where $t$ is the cosmic time.
In all these cases, the conformal time
is related to the expansion parameter through
\begin{equation}
\label{epsilon}
a \eta H(\eta)  = -(1+\epsilon),
\end{equation}
where $H = \mathcal{H}/a$ is the Hubble parameter, and
$\epsilon = 0$ for de Sitter inflation, $\epsilon = \epsilon_1$ for
slow-roll inflation, $\epsilon = 1/q$ for power-law inflation with $q \gg 1$.
Defining the complex function
\begin{equation}
\label{BBprime}
m_k = \eta_e \, \frac{\B_{\kk}^*(\eta_e) \cdot \B_{\kk}'(\eta_e)}{|\B_{\kk}(\eta_e)|^2} \, ,
\end{equation}
then, Eq.~(\ref{BB3}) becomes
\begin{eqnarray}
\label{BB4}
\B_{\kk}(\eta) \!\!& = &\!\! -[m_k - 2(1+\epsilon) - 1] \, \B_{\kk}(\eta_e) \left(\frac{a_e}{a}\right)^{\!2} \nonumber \\
\!\!& + &\!\! [m_k - 2(1+\epsilon)] \, \B_{\kk}(\eta_e) \left(\frac{a_e}{a}\right)^{\!2} \! \left(\frac{\eta}{\eta_e}\right) \! .
\end{eqnarray}
Let us observe that if the magnetic field evolves
as a simple power of the conformal time during inflation, or more generally at the end of it,
then the quantity $m_k$ is simply the exponent of that power,
\begin{equation}
\label{Betam}
\B_{\kk}(\eta) \propto \eta^{m_k}.
\end{equation}
Accordingly, looking at Eq.~(\ref{BB4}), we see that if the magnetic field during
inflation evolves adiabatically, $m_k = 2(1+\epsilon)$, the evolution after inflation is still adiabatic, and we reobtain the results in Sec.~II.
If during inflation, instead, the magnetic field evolves as a power of the conformal time
with exponent
\begin{equation}
\label{BB6bis}
m_k = 2(1+\epsilon) + \delta m_k,
\end{equation}
then the evolution after inflation is adiabatic for
times $|\eta| \lesssim |\eta_\star|$ and superadiabatic for times
$|\eta_\star| \lesssim |\eta| \lesssim |\eta_\downarrow|$, where
\begin{equation}
\label{BB6}
|\eta_\star| = \frac{1}{|\delta m_k|} \, |\eta_e|.
\end{equation}
Accordingly, if the evolution of magnetic fields during inflation is such that
$|\delta m_k|$ is a nonzero order-one function, then the evolution of superhorizon magnetic fields is superadiabatic up to their reentering the horizon.
If, instead, $|\delta m_k|$ is a small quantity of the same order of magnitude of $|k\eta_e|$ or smaller,
then superhorizon magnetic fields evolve adiabatically after inflation.

In particular, if one assumes that during inflation and on superhorizon scales
\begin{equation}
\label{Tsagasm}
\B_{\kk}(\eta) \propto \eta^m,
\end{equation}
with $m$ being a real constant such that $0 \leq m < 2$, then the postinflationary evolution is superadiabatic (unless $m$ is very near to 2), in agreement with the result of~\cite{Tsagas2}.

Finally, let us observe that, taking into account Eq.~(\ref{x6}), it easy to see that $\delta m_k$ is related to the quantity $g(n_\kk,\Omega_k,k\eta)$ defined in Eq.~(\ref{x6bis}) through the relation
\begin{equation}
\label{BB5}
\delta m_k = - ik\eta_e g(n_\kk,\Omega_k,k\eta_e).
\end{equation}
Taking into account Eqs.~(\ref{BB4}), (\ref{BB6bis}), and (\ref{BB5}),
one easily recovers all the results obtained in
Sec.~IIIb.

\subsection{IVb. The Ratra model}

The Ratra model~\cite{Ratra} is described by a nonconformal-invariant electromagnetic Lagrangian of the form
\begin{equation}
\label{RR1}
{\mathcal L}_{\rm em} = -\frac14 f(\phi) F_{\mu\nu} F^{\mu\nu},
\end{equation}
where the function $f(\phi)$ kinematically couples the inflaton $\phi$, the scalar field
responsible for inflation, to the photon. In this model,
the coupling function is a power-law function of the conformal time,
\begin{equation}
\label{RR2}
f(\eta) =
\left\{ \begin{array}{lll}
  (\eta_e/\eta)^{2p} \, ,  & \eta \leq \eta_e, \\
  1,                       & \eta > \eta_e,
  \end{array}
  \right.
\end{equation}
where $p \leq 0$ in order to avoid the strong-coupling problem.
\footnote{The case $p > 0$ can be worked out in a similar manner provided that a slightly different
form of $f(\eta)$ is considered in order to avoid the strong-coupling problem~\cite{Campanelli1}.
However, in order to avoid inessential complications, we will analyze this case elsewhere~\cite{Marrone}.}
In the case of pure de Sitter inflation, the Bogolubov coefficients
in Eq.~(\ref{BD}) are easily found in the Ratra model~\cite{Campanelli1}.
From these, it follows that the number of created photons $n_\kk$ is vanishing
for $p=0$ and $p=-1/2$, and it is approximatively given by
\begin{equation}
\label{RR3}
n_\kk \simeq n_p (-k\eta_e)^{-2\nu_p-1}
\end{equation}
at the lowest order in $-k\eta_e \ll 1$. Here, $\nu_p = |p+1/2|$,
\begin{equation}
\label{RR4}
n_p =
\left\{ \begin{array}{llll}
  [2^{\nu_p}(1-2\nu_p)\Gamma(\nu_p)]^2/32\pi,            & \nu_p \neq 0,
  \\
  \{\pi^2 + 4[2 + \gamma + \ln(-k\eta_e/2)]^2\}/32\pi, & \nu_p = 0,
  \end{array}
  \right.
\end{equation}
$\gamma$ is the Euler-Mascheroni constant, and $\Gamma(x)$ is the gamma function.
Moreover, using the results of ~\cite{Campanelli1}, the $\Omega_k$ angle (the phase difference between the two Bogolubov coefficients) is easily found to be
\begin{equation}
\label{RR5}
\Omega_k \simeq \pm \pi - \Omega_p k\eta_e
\end{equation}
at the lowest order in $-k\eta_e$, where $\pm$ corresponds to $\nu_p \gtrless 1/2$, respectively, and
\begin{equation}
\label{Omegap}
\Omega_p = 2 \, \frac{1+2\nu_p}{1-2\nu_p} \, .
\end{equation}
From Eqs.~(\ref{RR3}) and (\ref{RR5}) it follows that $n_\kk \gg 1$ and
$n_\kk^{-1} \ll |k\eta_e|$ for all $\nu_p \neq 1/2$.
\footnote{For $\nu_p \neq 0$ this comes straightforwardly. In the case $\nu_p = 0$, we have
$6 \times 10^{-3} \lesssim n_\kk^{-1}/|k\eta_e| \lesssim 6 \times 10^{-1}$ if we use
the result in Sec.~V that $1 \times 10^{-27} \lesssim |k\eta_e| \lesssim 5 \times 10^{-2}$ for any
realistic magnetogenesis scenario.}
Inserting Eqs.~(\ref{RR5}) in Eq.~(\ref{Io1}), and expanding in terms of $n_\kk$ and (afterwards) in terms of $-k\eta_e$, we find, at the lowest order
\begin{equation}
\label{RRx}
a^2 B_k(\eta) \simeq \frac{\sqrt{2}k^2}{\pi} \, \sqrt{n_\kk} \, | \mbox{$\frac12$} \Omega_p k\eta_e  + k\eta|.
\end{equation}
The above asymptotic expansion can also be obtained by inserting the exact expressions of the Bogolubov coefficients derived in~\cite{Campanelli1} in Eq.~(\ref{Io1}) and expanding in terms of $-k\eta_e$.

Since $\Omega_p$ is an order-one factor (excluding a very narrow region around $\nu_p = 1/2$,
where $\Omega_p$ diverges), Eq.~(\ref{RRx}) implies that, after inflation, superhorizon magnetic fields in the Ratra model rapidly approach a state B1 [compare  Eq.~(\ref{RRx}) with the first equation in Eq.~(\ref{Io3})].
(The state of such fields is an example of what we call a state CB1,
which will be formally introduced and discussed in Sec.~V.)
We conclude that, in the Ratra model, large-scale postinflationary magnetic fields
evolve superadiabatically up to their reentering the horizon.

The fact that superhorizon, inflation-produced magnetic fields in the Ratra model
evolve superadiabatically can be derived also in the following way.
Taking into account Eq.~(\ref{BB5}) and using the results of~\cite{Campanelli1},
we find
\begin{equation}
\label{RR6}
\delta m_k = \frac12 - \nu_p + \varepsilon_k,
\end{equation}
where $\varepsilon_k$ is a function of $-k\eta_e$ such that $\varepsilon_k \rightarrow 0$
for $-k\eta_e \rightarrow 0$. At the lowest order in $-k\eta_e$, we find
\begin{equation}
\label{RR7}
\varepsilon_k \simeq
\left\{ \begin{array}{llll}
  \frac{1}{2(\nu_p - 1)} \, (-k\eta_e)^2,                                           & \nu_p > 1,
  \\
  \left[i\pi/2 - \gamma - \ln(-k\eta_e/2)\right] (-k\eta_e)^2,                      & \nu_p = 1,
  \\
  \frac{2\pi[i-\cot(\pi\nu_p)]}{[2^{\nu_p}\Gamma(\nu_p)]^2} \, (-k\eta_e)^{2\nu_p}, & 0 < \nu_p < 1,
  \\
  \left[-i\pi/2 + \gamma + \ln(-k\eta_e/2)\right]^{-1},                             & \nu_p = 0.
  \end{array}
  \right.
\end{equation}
Since $\delta m_k$ is an order one constant in the limit $-k\eta_e \rightarrow 0$
(excluding the case $\nu_p = 1/2$, which, however corresponds to the case where there is no production of photons) it follows, according to the discussion in Sec.~IVa, that the magnetic field evolves superadiabatically.
\footnote{In the Ratra model discussed in~\cite{Campanelli1}, the magnetic field $\B_{\kk}$ and its first
derivative are not in general continuous functions of the conformal time at $\eta = \eta_e$.
The relevant quantities are, however, the rescaled magnetic field $\boldsymbol{\Phi}_{\kk} = \sqrt{f} \, \B_{\kk}$ and its first derivative which, instead, are continuous functions.
As it is easy to check by using the results of~\cite{Campanelli1}, the rescaled magnetic field evolves
as in Eq.~(\ref{Betam}) on superhorizon scales, $\boldsymbol{\Phi}_{\kk}(\eta) \propto \eta^{m_k}$, with $m_k = 2 + \delta m_k$ and $\delta m_k$ given by Eq.~(\ref{RR6}).
After inflation, the quantity $\boldsymbol{\Phi}_{\kk}$
coincides with $\B_{\kk}$ since $f(\eta) = 1$ for $\eta \geq \eta_e$. Consequently, the magnetic field in the Ratra model evolves after inflation according to Eq.~(\ref{BB4}) with $\epsilon = 0$,
$m_k = 2 + \delta m_k$, and $\delta m_k$ given by Eq.~(\ref{RR6}).}

Particularly interesting is the case $\nu_p = 3/2$, which corresponds to $p = -2$.
In this case, in fact, the particle number is proportional to  $n_\kk \propto (-k\eta_e)^{-4}$,
so that the actual magnetic field is scaling invariant [see Eqs.~(\ref{D4})].
Moreover, in this case, $n_\kk$, $\Omega_k$, and $\delta m_k$
have the simple expressions
\begin{eqnarray}
\label{RR8}
\!\!\!\!\!\!\!\! && n_\kk = \mbox{$\frac14$} (-k\eta_e)^{-4}, \\
\label{RR9}
\!\!\!\!\!\!\!\! &&\Omega_k = \pi + 2 k\eta_e + \arctan \! \left[\frac{2k\eta_e}{1- 2(k\eta_e)^2}\right] \! , \\ %\; (\mbox{mod}(-\pi,\pi]), \\
\label{RR10}
\!\!\!\!\!\!\!\! &&\delta m_k = -ik\eta_e - (1 + ik\eta_e)^{-1},
\end{eqnarray}
respectively, valid for all wave numbers $k$.

In the right panel of Fig.~1, the curve referred to as ``Ratra'' shows the magnetic flux spectrum
$a^2 B_k(\eta)$ normalized to its expression in the
case of pure Maxwell theory, $k^2/\sqrt{2}\pi$, as a function of $k\eta$
in the scaling-invariant case, where $n_\kk$ and $\Omega_k$ are given by Eqs.~(\ref{RR8}) and (\ref{RR9}), respectively. As is clear from the figure, and as we have discussed above,
superhorizon ($k\eta \ll 1$) magnetic fields
in the Ratra model evolve superadiabatically as $a^2 B_k(\eta) \propto \eta$
up to their reentering the horizon.

\section{V. Initial magnetic state}

The discussion in Sec.~III on the evolution of postinflationary superhorizon magnetic fields
has been as general as possible. We have seen that the evolution after inflation of a given
magnetic mode with wave number $k$, crucially depends on three parameters, namely
$k\eta_e$, $n_\kk$, and $\Omega_k$. It turns out that $\Omega_k$ cannot be constrained
by present cosmological observations, while $k\eta_e$ and $n_\kk$ are directly
connected to cosmological observables, such as the scale of inflation $M$ and the reheat
temperature $T_{\rm RH} < M$ on the one hand, and the actual magnetic field intensity on the scale $\lambda = 1/k$
on the other hand.

To see this, let us first observe that
the expansion parameter at the end of inflation is given by
\begin{equation}
\label{DD0}
a_e^3 = \frac{\pi^2}{30} \, g_{*S,0} \, \frac{g_{*,{\rm RH}}}{g_{*S,{\rm RH}}}
\frac{T_0^3 \, T_{\rm RH}}{M^4},
\end{equation}
where $T_0 \simeq 2.35 \times 10^{-4} \eV$~\cite{Fixsen} is the actual temperature,
$g_{*S,0} = g_{*S}(T_0)$, $g_{*S,{\rm RH}} = g_{*S}(T_{\rm RH})$, and
$g_{*,{\rm RH}} = g_{*}(T_{\rm RH})$. Here, $g_{*}(T)$ and $g_{*S}(T)$ are the effective
number of degrees and entropy degrees of freedom at the temperature $T$, respectively~\cite{Kolb-Turner}.
For temperatures above $T \sim 0.1\MeV$, the quantities $g_{*}(T)$ and $g_{*S}(T)$ can be considered equal,
while below $T \sim 0.1\MeV$ these quantities equal the corresponding quantities
evaluated at the present time, $g_{*,0} = 2 + (21/11) (4/11)^{1/3}$ %\simeq 3.36
and $g_{*S,0} = 43/11$~\cite{Kolb-Turner}.
In obtaining Eq.~(\ref{DD0}), we used the following facts. First,
the energy scale of inflation is defined by $\rho(\eta_e) = M^4$,
where $\rho$ is the energy density of the Universe.
Second, during reheating $\rho$ scales approximatively as $\rho \propto a^{-3}$.
Third, the reheat temperature is defined by $\rho(\eta_{\rm RH}) = (\pi^2/30) \, g_{*,\rm{RH}} \, T_{\rm RH}^4$,
where $\eta_{\rm RH}$ is the time at the end of reheating. Finally,
from the end of reheating until today, the expansion parameter is related
to the temperature $T$ through $a(T) \propto g_{*S}^{-1/3} T^{-1}$~\cite{Kolb-Turner}.

Using Eqs.~(\ref{epsilon}) and (\ref{DD0}), and taking into account the
fact that the Hubble parameter $H$ is related to the energy density of the Universe through the Friedmann equation $H^2 = 8\pi \rho/3\mPl^2$, with $\mPl$ being the Planck mass, we get
\begin{equation}
\label{DD1}
-k\eta_e \simeq 1 \times 10^{-23} M_{16}^{-2/3} T_{16}^{-1/3} \lambda_{\rm Mpc}^{-1},
\end{equation}
where $M_{16} = M/10^{16} \GeV$, $T_{16} = T_{\rm RH}/10^{16} \GeV$,
and $\lambda_{\rm Mpc} = \lambda/\Mpc$. In order to be consistent with
cosmic microwave background observations, the scale of inflation $M$, which is directly
related to the amplitude of the primordial tensor perturbations, has to be, roughly speaking,
below $10^{16} \GeV$~\cite{Kolb-Turner}.
The minimum value for the reheat temperature is around $4.7 \MeV$~\cite{Mangano-Miele}.
This constraint, which comes from the analysis of cosmic microwave background radiation data,
assumes a scale of inflation greater than about $43 \MeV$.

The actual scale of astrophysics interest for cosmic magnetic fields ranges from
the minimum scale for having a successful galactic dynamo~\cite{Davis},
\footnote{Large-scale galactic dynamo could, in principle, explain the
presence of galactic magnetic fields if a sufficiently strong seed field were present
prior to galaxy formation. However, galactic dynamos leave substantially unanswered
the question of the presence of strong magnetic fields in clusters of galaxies and cosmic voids.}
\begin{equation}
\label{dynamo1}
\lambda_{\rm dyn, min} \sim 100 \pc,
\end{equation}
to the present horizon, $\eta_0 \simeq 14185 \Mpc$.
\footnote{The present horizon is $\eta_0 = H_0^{-1} \! \int_0^{\infty} \! dz E^{-1}(z)$,
where $H_0$ is the Hubble constant, $z$ is the redshift, and
$E(z) = \sqrt{\Omega_r (1+z)^4 + \Omega_m (1+z)^3 + \Omega_\Lambda}$.
Here, $\Omega_r$, $\Omega_m$, and $\Omega_\Lambda$ are the radiation, matter, and cosmological constant density parameters,
respectively, which in a spatially flat Universe satisfy the relation $\Omega_r + \Omega_m + \Omega_\Lambda = 1$.
Using the fact that $\Omega_r h^2 = 4.31 \times 10^{-5}$~\cite{Kolb-Turner}, where $h$ is the normalized
Hubble constant $H_0 = 100 h \km \s^{-1} \Mpc^{-1}$,
and the Planck results~\cite{Planck2015} $\Omega_m = 0.308$ and $h = 0.678$, we find the value of $\eta_0$ given in the text.}
Accordingly, and as we have supposed in the previous sections, the quantity $|k\eta_e|$ in Eq.~(\ref{DD1})
is much smaller than one. In fact, its minimum and maximum values are, respectively,
$|k\eta_e|_{\rm min} \simeq 1 \times 10^{-27}$, corresponding to take an instantaneous reheating
\footnote{Instantaneous reheating refers to the ideal case where after inflation the
Universe enters directly in the radiation-dominated era.
In this case, equating the energy density of radiation at the beginning of radiation era,
which is the same as the energy density at the end of reheating, to the energy
density at the end of inflation,
we get $T_{\rm RH} = [30/(\pi^2 g_{*,\rm{RH}})]^{1/4} M$.
Taking $g_{*,\rm{RH}} = 427/4$, referring to the massless degrees of
freedom of the standard model of particle physics above the electroweak scale,
we find $T_{\rm RH} \simeq 0.4 M$.}
with $M = 10^{16} \GeV$, and $\lambda = \eta_0$, and $|k\eta_e|_{\rm max} \simeq 5\times 10^{-2}$,
corresponding to take $M = 43 \MeV$, $T_{\rm RH} = 4.7\MeV$, and $\lambda = 100\pc$.

Let us now show that the particle number $n_\kk$ is a quantity much greater than one, as we assumed in the previous sections.
To this end, let us rewrite Eq.~(\ref{D4}) as
\begin{equation}
\label{DD2}
B_k(\eta_0) \simeq 2 \times 10^{-13} \zeta_k \left( \frac{n_\kk}{10^{89}} \right)^{\!1/2} \lambda_{\rm Mpc}^{-2} \G.
\end{equation}
In order to explain directly (i.e., without invoking any galactic dynamo)
the presence of large-scale magnetic fields in galaxies and clusters of galaxies
it suffices to have a seed magnetic field with correlation length and strength
in the ranges~\cite{Campanelli,Campanelli2}
\begin{eqnarray}
\label{range1}
&& few \times \Mpc \lesssim \lambda \lesssim \eta_0, \\
\label{range2}
&& 10^{-13} \G \lesssim B_k(\eta_0) \lesssim few \times 10^{-12} \G.
\end{eqnarray}
From Eq.~(\ref{DD2}), then, we see that a particle number in the range
$10^{89} \lesssim n_\kk \lesssim 10^{99}$
is needed to directly explain cosmic magnetic fields.
In general, allowing a very efficient galactic dynamo and then a very weak seed magnetic field
results in a needed particle number much less than the above, but still much greater than unity.
To see this, let us rewrite Eq.~(\ref{DD2}) as
\begin{equation}
\label{DD3}
n_\kk \simeq 3 \times 10^{38} \zeta_k^{-2} \lambda_{\rm 100 pc}^4 B_{30}^{2},
\end{equation}
where $\lambda_{\rm 100 pc} = \lambda/100\pc$ and $B_{30} = B_k(\eta_0)/10^{-30} \G$. The minimum photon number
is $n_{\kk,\rm min} \simeq 1 \times 10^{39}$, corresponding to take $\lambda = 100\pc$,
$\zeta_k = \zeta_{k,\rm max} = \sqrt{2}/\pi$, and the very ``optimistic'' lower bound
on a seed magnetic field that a galactic dynamo can amplify up to the observed values~\cite{Davis},
\begin{equation}
\label{dynamo2}
B_{\rm dyn,min}(\eta_0) \sim 10^{-30} \G.
\end{equation}
It is worth noticing, however, that the results in~\cite{Davis} have been strongly criticized
in the literature and the minimum value of $10^{-30} \G$ for a seed magnetic field
seems to be unrealistically small (see, e.g.,~\cite{Widrow}). In any case, the
exact value of $B_{\rm dyn,min}(\eta_0)$ is not important for our discussion.
\footnote{It is worth noticing that another possible amplification mechanism of seed magnetic fields is the so-called small-scale dynamo which, in contrast to the large-scale one,
can work both in galaxies and in the intracluster medium
(for a review on large- and small-scale dynamos, see~\cite{dynamo}).
Therefore, a small-scale dynamo could, at least in principle, explain the presence of cosmic magnetic fields in galaxies and galaxy clusters if a seed field were present before large-scale structure formation.
However, as in the case of large-scale dynamo, a small-scale dynamo cannot account for the presence of
large-scale magnetic fields in cosmic voids.}

We can now show that, in any realistic model of inflationary magnetogenesis, the quantity
$|k\eta_e|$ is always much greater than $n_\kk^{-1}$. In fact, taking into account the
results~(\ref{DD1}) and (\ref{DD3}), we have
\begin{equation}
\label{DD4}
\frac{|k\eta_e|}{n_\kk^{-1}} \simeq 3 \times 10^{14} \zeta_k^{-2} M_{16}^{-2/3} T_{16}^{-1/3}
\lambda_{\rm 100 pc}^3 B_{30}^{2},
\end{equation}
from which it follows that the minimum value of $(|k\eta_e|/n_\kk^{-1})$, corresponding to take an
instantaneous reheating with $M = 10^{16} \GeV$, $\lambda = 100\pc$, $\zeta_k = \zeta_{k,\rm max}$,
and $B_k(\eta_0) = 10^{-30} \G$, is $(|k\eta_e|/n_\kk^{-1})_{\rm min} \simeq 2 \times 10^{15}$.

Focusing our discussion on superhorizon magnetic fields that may eventually explain cosmic magnetization,
we conclude that, whatever is the mechanism responsible for their generation,
they start their evolution after inflation in a state characterized by
\begin{equation}
\label{Leo}
n_\kk^{-1} \ll |k\eta_e| \ll 1.
\end{equation}
Looking at Eqs.~(\ref{Q1}) and (\ref{Q2}), then, the possible initial magnetic states are:
A1, B1, B2, C1 (which correspond to cases shown in the right
panel of Fig.~1). For the sake of completeness, let us observe that
just another possible initial state exists.
It is defined by
\begin{equation}
\label{post1}
\mbox{case CB1} \!: \;\; n_\kk^{-1} \ll |k\eta_e| \sim |\Omega_k \mp \pi| \ll 1.
\end{equation}
This case, however, does not differ qualitatively and, roughly speaking, quantitatively by the case where the initial state is described by either the case C1 or the case B1. In fact, let us write
\begin{equation}
\label{post2}
\Omega_k \mp \pi = -\omega_k k\eta_e,
\end{equation}
where $\omega_k$ is an order-one function of $k$.
Inserting Eq.~(\ref{post1}) in Eq.~(\ref{Io1}), and expanding in terms of $n_\kk$ and (afterwards) in terms of $-k\eta_e$, we get
\begin{equation}
\label{post3}
a^2 B_k(\eta) \simeq \frac{\sqrt{2}k^2}{\pi} \, \sqrt{n_\kk} \, | \mbox{$\frac12$} \omega_k k\eta_e  + k\eta| \, , \;\, \mbox{CB1}.
\end{equation}
For $\eta = |\eta_e|$, the above expression becomes
\begin{eqnarray}
\label{postpost}
\frac{a^2 B_k(|\eta_e|)}{k^2/\sqrt{2}\pi} \!\!& \simeq &\!\! |1-2/\omega_k| |\Omega_k \mp \pi| \sqrt{n_\kk}
\nonumber \\
                  \!\!& \simeq &\!\! 2 |1-\omega_k/2| \sqrt{n_\kk} \, |k\eta_e|.
\end{eqnarray}
%
%$a^2 B_k(|\eta_e|) = (k^2/\sqrt{2}\pi) |1-2/\omega_k| |\Omega_k \mp \pi| \sqrt{n_\kk}$ or,
%equivalently, $a^2 B_k(|\eta_e|) \simeq (k^2/\sqrt{2}\pi) |1-\omega_k/2| 2\sqrt{n_\kk} |k\eta_e|$.
Comparing these expressions with the first two equations of Eq.~(\ref{Io3}) and neglecting order-one factors, we see that the magnetic field in an initial state CB1 can be viewed as being in an initial state C1 or, equivalently, B1, as anticipated. At later times, instead, the
case CB1 converges to the case B1. In fact, Eq.~(\ref{post3}) reduces to the first equation of Eq.~(\ref{Io3}) in the limit $\eta \gg |\eta_e|$. The time when this convergence happens can be estimated as
$\eta_\star \simeq |1-2/\omega_k| \eta_{\tiny \mbox{C1} \rightarrow \mbox{B1}} = |1-\omega_k/2| |\eta_e|$, which agrees with the fact that a magnetic field in the initial state CB1 can be viewed, approximatively, as
being in an initial state either C1 or B1.

Summarizing, we have found that superhorizon magnetic fields start their evolution after inflation in one of the following states:

\vspace{0.15cm}

\parbox[t]{0.3cm}{$\bullet$} \parbox[t]{7.4cm}{{A1, in which case the evolution is adiabatic;}}

\parbox[t]{0.3cm}{$\bullet$} \parbox[t]{7.4cm}{{B1 or B2, in which cases the evolution is super-adiabatic;}}

\vspace{0.08cm}

\parbox[t]{0.3cm}{$\bullet$} \parbox[t]{7.4cm}{{C1 or CB1, in which case there is an adiabatic evolution followed by a superadiabatic evolution.}}

\section{VI. Discussion}

Taking into account the results of Sec.~III and those of the previous section, we can
relate the magnetic field at the end of inflation to its present value through
\begin{equation}
\label{DD5}
B_k(\eta_e) = \left(\frac{a_0}{a_e}\right)^{\!2}  \! B_k(\eta_0) \,  |k\eta_\star|,
\end{equation}
where $|\eta_e| \lesssim |\eta_\star| \lesssim |\eta_\downarrow|$ is the time when
superadiabatic evolution starts, and
\begin{equation}
\label{DD6}
|k\eta_\star| \simeq
\left\{ \begin{array}{lll}
  1,                           & \mbox{A1},
  \\
  |k \eta_e|,                  & \mbox{B1, B2},
  \\
  |1-\omega_k/2| |k \eta_e|,   & \mbox{CB1},
  \\
  \frac12 |\Omega_k \mp \pi|,  & \mbox{C1}.
  \end{array}
  \right.
\end{equation}
Equations~(\ref{DD5}) show that an amplification of a factor $1/|k\eta_\star|$
%(which is maximum for the cases B1 and B2)
of the actual magnetic field spectrum
occurs with respect to the case where causality is not imposed on the
evolution of superhorizon magnetic fields after inflation,
to wit, with respect to the case where postinflationary magnetic fields
are assumed to evolve adiabatically on superhorizon scales.

\subsection{VIa. Comparison with Tsagas results}

Let us now compare our results with those obtained by Tsagas.
As we have shown in Sec.~IVa, the case analyzed in~\cite{Tsagas2} corresponds to our
cases B1 and B2, where the magnetic evolution on superhorizon scales is superadiabatic.

Taking into account Eqs.~(\ref{DD6}) and (\ref{DD0}),
we can rewrite Eq.~(\ref{DD5}) (cases B1 and B2) as
\begin{equation}
\label{DD7}
B_k(\eta_e) = c_0 \, \frac{\mPl M^2 k}{T_{\rm RH} T_0^3} \, B_k(\eta_0),
\end{equation}
where $c_0 = 15(3/2\pi^5)^{1/2}(1+\epsilon)g_{*S,0}^{-1} \simeq 0.3$ %0.27
and we assumed, as in ~\cite{Tsagas2}, a de Sitter inflation.

In radiation-dominated era, the temperature when a given mode $k$ crosses
inside the horizon is given by
\footnote{The condition $k \eta_\downarrow = 1$ that a mode $k$ crosses inside the horizon
can be re-written as $2t_\downarrow k/a_\downarrow = 1$, where $t_\downarrow$ and $a_\downarrow$
are the cosmic time and the expansion parameter at the time of crossing.
Here, we used the fact that in radiation-dominated era $\eta = 2t/a$. Since $a = (g_{*S,0}/g_{*S})^{1/3} T_0/T$,
and since $t = (45/16\pi^3 g_{*})^{1/2} \mPl/T^2$~\cite{Kolb-Turner} in radiation-dominated era, we recover Eq.~(\ref{Tdownr}).}
\begin{equation}
\label{Tdownr}
T_{\downarrow} = \kappa_r \, \frac{\mPl k}{T_0} \, ,
\end{equation}
where $\kappa_r = (45/4\pi^3 g_{*,\downarrow})^{1/2} (g_{*S,\downarrow}/g_{*S,0})^{1/3}$,
with $g_{*,\downarrow} = g_{*}(T_{\downarrow})$ and $g_{*S,\downarrow} = g_{*S}(T_{\downarrow})$.
In matter-dominated era, instead, we have
\footnote{The condition that a mode $k$ crosses inside the horizon
in matter-dominated era can be re-written as $3t_\downarrow k/a_\downarrow = 1$, where
we used the fact that $\eta = 3t/a$ in this era.
Since $t = (5/\pi^3 g_{*,0})^{1/2} (g_{*S,0}/g_{*S})^{1/3} \mPl/T_{\rm eq}^{1/2}T^{3/2}$~\cite{Kolb-Turner}
in matter-dominated era, we recover Eq.~(\ref{Tdownm}).}
\begin{equation}
\label{Tdownm}
T_{\downarrow}^{1/2} T_{\rm eq}^{1/2} = \kappa_m \, \frac{\mPl k}{T_0} \, ,
\end{equation}
where $T_{\rm eq} = (\Omega_m/\Omega_r) \, T_0 \simeq 0.8 \eV$~\cite{Kolb-Turner,Planck2015} is
the temperature at the radiation-matter transition, and $\kappa_m = (45/\pi^3 g_{*0})^{1/2}$.

In radiation-dominated era, then, Eq.~(\ref{DD7}) can be re-written as
\begin{equation}
\label{DD10}
B_k(\eta_e) = c_r  \, \frac{M^2 T_{\downarrow}}{T_{\rm RH} T_0^2} \,  B_k(\eta_0),
\end{equation}
where $c_r = c_0/\kappa_r
%= (30/\pi^2)^{1/2} (1+\epsilon) g_{*S,0}^{-2/3} \, g_{*S,\downarrow}^{-1/3} \, g_{*,\downarrow}^{1/2}
\simeq 0.7 g_{*S,\downarrow}^{-1/3} \, g_{*,\downarrow}^{1/2}$ is slowly increasing function of $T_{\downarrow}$ of order one
such that $0.8 \lesssim c_r \lesssim 1.5$.
Assuming $\lambda = 1/k \gtrsim 100 \pc$, we have $T_\downarrow(\lambda) \lesssim T_\downarrow (100\pc) \simeq 0.9 \MeV$, %0.85 \MeV
which implies $0.8 \lesssim c_r \lesssim 1.0$
(here, we used the fact that $g_*(T) = 43/4$ for $m_e \lesssim T \lesssim m_\mu$,
with $m_e \simeq 0.5 \MeV$ and $m_\mu \simeq 106 \MeV$ being the mass of the electron and muon,
respectively).

In matter-dominated era, instead, Eq.~(\ref{DD7}) reads
\begin{equation}
\label{DD11}
B_k(\eta_e) = c_m \, \frac{M^2 T_{\downarrow}^{1/2} T_{\rm eq}^{1/2}}{T_{\rm RH} T_0^2} \, B_k(\eta_0),
\end{equation}
where $c_m = c_0/\kappa_m
%= (15/2\pi^2)^{1/2} (1+\epsilon) g_{*S,0}^{-1} \, g_{*,0}^{1/2}
\simeq 0.4$. %0.41

Equations~(\ref{DD10}) and (\ref{DD11}) coincide, apart from numerical factors of order one, with the results of~\cite{Tsagas2}.

\subsection{VIb. Implications for the Ratra model}

Any model of inflationary magnetogenesis can be trustful if the
inflation-produced (electro)magnetic field does not appreciably
backreact on the dynamics of the Universe.
After reheating, electric fields inside the horizon are washed out by the high conductivity of the primeval plasma, while subhorizon magnetic fields evolve adiabatically (neglecting any effect of magnetohydrodynamic turbulence).
It is a well-known result that, if such magnetic fields have to explain the observed cosmic magnetic fields,
then their energy after inflation is always negligible with respect to that of the Universe.

In order to avoid the backreaction problem during inflation,
the VEV of the electromagnetic energy must be subdominant with respect to the
energy density of inflation,
\begin{equation}
\label{I1}
\langle 0| \rho_{\rm em}(\eta,\x)|0 \rangle \ll \rho_{\rm inf}(\eta).
\end{equation}
Assuming, as in Sec.~IVb and for the sake of simplicity, a de Sitter inflation, we have
$\rho_{\rm inf}(\eta) = M^4 = 3\mPl^2H_{\rm dS}^2/8$, where
$H_{\rm dS} \ll M$ is the Hubble parameter during de Sitter inflation. The electromagnetic energy density is,
instead,
\begin{equation}
\label{I2}
\rho_{\rm em}(\eta,\x) = \rho_E(\eta,\x) + \rho_B(\eta,\x),
\end{equation}
where
\begin{eqnarray}
\label{energies}
\rho_E = \frac12 f(\eta) \E^2, \;\;\; \rho_B = \frac12 f(\eta) \B^2
\end{eqnarray}
are the electric and magnetic energy densities, respectively,
and $a^2 \E = -\dot{\A}$ is the electric field.
Using Eq.~(\ref{F1}), we find
\begin{equation}
\label{www}
\langle 0| \rho_{\rm em}|0\rangle = \int_0^{\infty} \! \frac{dk}{k} \, \rho_{{\rm em},k}(\eta),
\end{equation}
where $\rho_{{\rm em},k}(\eta) = \rho_{E,k}(\eta) + \rho_{B,k}(\eta)$
is the electromagnetic energy spectrum and
\begin{eqnarray}
\label{I3}
&& \rho_{E,k}(\eta) =  f(\eta) \, \frac{k^2}{2\pi^2 a^4} \sum_{\alpha=1,2} |\dot{A}_{k,\alpha}(\eta)|^2, \\
\label{I4}
&& \rho_{B,k}(\eta) = f(\eta) \, \frac{k^4}{2\pi^2 a^4} \sum_{\alpha=1,2} |A_{k,\alpha}(\eta)|^2
\end{eqnarray}
are the electric and magnetic energy spectra, respectively.

For large wave numbers (subhorizon modes), $-k\eta \gg 1$, $\rho_{{\rm em},k}(\eta)$ reduces to
the electromagnetic energy spectrum in the free Maxwell theory (the case $p=0$),
$\rho_{{\rm em},k}(\eta) = (-k\eta)^4 H_{\rm dS}^4/2\pi^2$,
which is ultraviolet divergent. Such a kind of divergence, however, can be cured
by renormalization~\cite{Birrell-Davies,Parker-Toms}.
Indeed, it can be shown that, after renormalization, the
electromagnetic energy on subhorizon scales does not appreciably backreact
on inflation~\cite{Campanelli1}.

For $p \neq 0$ and $-k\eta \ll 1$ (superhorizon scales), instead, the electromagnetic spectrum is dominated by the electric component~\cite{Campanelli1,Marrone},
\begin{equation}
\label{I5}
\rho_{{\rm em},k}(\eta) \sim \rho_{E,k}(\eta) \sim (-k\eta)^{2(p+2)} H_{\rm dS}^4,
\end{equation}
where, hereafter, we neglect numerical factor of order unity.
From the above equation, we find that backreaction on (de Sitter) inflation is
negligible if $-2 \leq p < 0$ or, equivalently, $0 \leq \nu_p \leq 3/2$.

The magnetic energy spectrum on superhorizon scales is
\begin{equation}
\label{I6}
\rho_{B,k}(\eta) \sim (-k\eta)^{5-2\nu_p} H_{\rm dS}^4.
\end{equation}
Since $f(\eta_e) = 1$, at the end of inflation we have $B_k(\eta_e) = \sqrt{2\rho_{B,k}(\eta_e)}$, so that
\begin{equation}
\label{I7}
B_k(\eta_e) \sim (-k\eta_e)^{\frac52 - \nu_p} H_{\rm dS}^2.
\end{equation}
According to the ``standard'' reasoning in the literature,
the inflation-produced magnetic field scales adiabatically
after reheating, since the highly conductive primeval plasma freezes it on all scales.
According to the Barrow-Tsagas causality arguments, instead,
only subhorizon magnetic modes are frozen into the plasma after reheating.
Superhorizon modes, instead, are superadiabatically amplified
by a factor $(-k\eta_e)^{-1}$ with respect to the previous case
[see Eqs.~(\ref{DD5}) and (\ref{DD6}), case CB1].
Accordingly, we have
\begin{equation}
\label{I8}
B_k(\eta_0) \sim
\left\{ \begin{array}{lll}
  (a_e/a_0)^2 (-k\eta_e)^{\frac52 - \nu_p} H_{\rm dS}^2,  & \mbox{\it \small without causality,}
  \\
  (a_e/a_0)^2 (-k\eta_e)^{\frac32 - \nu_p} H_{\rm dS}^2,  & \mbox{\it \small with causality,}
  \end{array}
  \right.
\end{equation}
where we have assumed an instantaneous reheating for the case of simplicity.
A scaling-invariant magnetic field is not possible today according to standard approach,
since it would correspond to the case $\nu_p = 5/2$,
which is excluded by the above backreaction arguments.
The maximum value for $B_k(\eta_0)$ is obtained for the maximum allowed value of $\nu_p$,
to wit, $\nu_p = 3/2$. In this case, instead, the actual magnetic field is scaling invariant
if one correctly takes into account the causality arguments. Thus, we have
\begin{equation}
\label{I9}
B_{k,{\rm max}}(\eta_0) \sim
\left\{ \begin{array}{lll}
  (-k\eta_e)/\eta_e^2,  & \mbox{\it \small without causality,}
  \\
  1/\eta_e^2,           & \mbox{\it \small with causality,}
  \end{array}
  \right.
\end{equation}
where we used the fact that $\eta  = -1/a H_{\rm dS}$ in de Sitter inflation.
Using Eq.~(\ref{DD1}) specialized to the case of instantaneous reheating, we finally get
\begin{equation}
\label{I10}
B_{k,{\rm max}}(\eta_0) \sim
\left\{ \begin{array}{lll}
  10^{-31} M_{16} \, \lambda_{\rm 100 pc}^{-1} \, \G,  & \mbox{\it \small without causality,}
  \\
  10^{-12} M_{16}^2 \, \G,                             & \mbox{\it \small with causality.}
  \end{array}
  \right.
\end{equation}
Taking into account Eqs.~(\ref{dynamo1}) and (\ref{dynamo2}), and
Eqs.~(\ref{range1}) and (\ref{range2}),
it follows, on the one hand, the standard result quoted in the literature that
the Ratra model cannot explain cosmic magnetism
and, on the other hand, the claim in~\cite{Campanelli1} that instead it can.

\section{VII. Conclusions}

The large-scale magnetic fields we observe today in galaxies, clusters of galaxies, and
cosmic voids are probably relics from inflation, and a possible explanation for them is
the creation of photons out from the vacuum through the Parker mechanism in
a putative (nonstandard), nonconformal-invariant theory of electrodynamics.

For long time since the first model of inflationary magnetogenesis by Turner and Widrow~\cite{Turner-Widrow}, it has been assumed that inflation-produced
magnetic fields remain frozen after reheating on superhorizon scales
(which are the scales of astrophysical interest for cosmic magnetic fields) due to the high
conductivity of the primeval plasma, so that they evolve adiabatically.
This assumption is, however, physically incorrect since it violates causality.
As first pointed out by Barrow and Tsagas~\cite{Barrow-Tsagas},
postinflationary electric currents are generated by microphysical processes
during reheating and, then, are vanishing on superhorizon scales. This, in turn, implies
a vanishing conductivity at scales larger than the Hubble radius after reheating.

The implications of this fact for cosmic magnetogenesis have been recently investigated by
Tsagas~\cite{Tsagas1,Tsagas2}. His results suggest that, in a spatially flat Friedmann-Robertson-Walker universe,
inflation-produced magnetic fields may be superadiabatically amplified after inflation
on superhorizon scales. Such an amplification may exist both in the conformal-invariant free Maxwell theory
and in a nonconformal-invariant electromagnetic theory where
magnetic fields evolve during inflation as power law of the conformal time.

Our results, if on the one hand show that magnetic fields in Maxwell theory starting in the Bunch-Davies vacuum evolve adiabatically, on the other hand indicate that
a superadiabatic evolution is, in principle, possible in the context of nonconformal-invariant theories of electrodynamics.

In particular, we have found that, irrespective of the particular underlaying electromagnetic theory, the evolution of the magnetic field spectrum $B_k(\eta)$ after inflation is ruled by the values of
three quantities: $k\eta_e$, $n_\kk$, and $\Omega_k$. Here, $k$ is the magnetic wave number, $\eta_e$ is the conformal time at the end of inflation, $n_\kk$ is the number density spectrum of photons produced out of the vacuum by inflation {\it via} the Parker mechanism and, finally, $\Omega_k$ is the
phase difference between the two Bogolubov coefficients which define the state of the electromagnetic mode $k$ at the end of inflation.

For generic models of inflation, we have found that the relation $n_\kk^{-1} \ll |k\eta_e| \ll 1$ holds
in any model of inflationary magnetogenesis which may eventually explain the presence of cosmic magnetic fields.
This, in turn, leaves open only three possibilities for the evolution of superhorizon magnetic fields after inflation:
($i$) $|\Omega_k \mp \pi| = \mathcal{O}(1)$, in which case the evolution is adiabatic, namely the magnetic flux spectrum $a^2 B_k(\eta)$ is constant in time;
($ii$) $|\Omega_k \mp \pi| \ll |k\eta_e|$, in which case the evolution is superadiabatic, in the sense that
the magnetic flux increases in time, in particular as $a^2B_k(\eta) \propto \eta$;
($iii$) $|k\eta_e| \ll |\Omega_k \mp \pi| \ll 1$ or $|k\eta_e| \sim |\Omega_k \mp \pi| \ll 1$, in which case
the evolution is adiabatic up the time $\eta_\star \sim |\Omega_k \mp \pi|/k$ and then superadiabatic afterwards,
with $a^2B_k(\eta) \propto \eta$.
In all cases, once a given magnetic mode reenters the horizon, it evolves adiabatically till today
since it remains frozen into the high-conductive primeval plasma.

We have applied our general results to two specific magnetogenesis scenarios.
First, we have found that the case studied by Tsagas, where $B_k(\eta) \propto |\eta|^m$ ($0 \leq m < 2$) during
inflation, belongs to the possibility ($ii$).
Second, we have found that postinflationary superhorizon magnetic fields evolve superadiabatically in the Ratra model~\cite{Ratra}. The consequences of the latter result reinforce our recent claim~\cite{Campanelli1} that the Ratra model can account for the presence of cosmic magnetic fields by evading both the backreaction and strong-coupling problems.

%**********************************   Bibliography   *****************************************%

\end{document}